\documentclass[preprint,12pt,number]{elsarticle}
\usepackage[english]{babel} % Consente l'uso caratteri accentati italiani
\usepackage{url}
\usepackage{graphicx}
\usepackage{array}
\usepackage{dcolumn}

\newcolumntype{d}[1]{D{.}{.}{#1}}

\journal{Journal of Electron Spectroscopy}
\begin{document}

\begin{frontmatter}

%\preprint{}

\title{Space-Charge effect in electron time-of-flight analyzer for high-energy photoemission spectroscopy} %Title of paper


\author{G. Greco\corref{cor1}}
    \ead{giorgia.greco@helmholtz-berlin.de}
\author{A. Verna\corref{}\fnref{cnism}}
\author{F. Offi\fnref{cnism}}
\author{G. Stefani\fnref{cnism}}
\cortext[cor1]{Corresponding author}
\address{Dipartimento di Scienze, Universit\`a Roma Tre, Via della Vasca Navale 84, I-00146 Rome, Italy}
\fntext[CNISM]{Also CNISM, Unit\`a Roma Tre, Via della Vasca Navale 84, I-00146 Rome, Italy.}
%\author{}
%\thanks{}
%\affiliation{}


\begin{abstract}
The space-charge effect, due to the instantaneous emission of many electrons  after the absorption of a single photons pulse, causes distortion in the photoelectron energy spectrum.
Two calculation methods have been applied to simulate the expansion during a free flight of clouds of mono- and bi-energetic electrons generated by a high energy pulse of light and their results have been compared.
%The main objective of this work is to test the accuracy of a widely used tool, such as SIMION\textsuperscript{\textregistered}, for predicting the energy distortion caused by the space-charge effect in any type of analyzer.
The accuracy of a widely used tool, such as SIMION\textsuperscript{\textregistered}, in predicting the energy distortion caused by the space-charge has been tested and the reliability of its results is verified. Finally we used SIMION\textsuperscript{\textregistered} to take into account the space-charge effects in the simulation of simple photoemission experiments with a time-of-flight analyzer.

\end{abstract}

%\maketitle %\maketitle must follow title, authors, abstract and \pacs

\date{September 4, 2016}
\end{frontmatter}
\section{Introduction}
New frontiers in solid state physics can be explored with the advent of the next generation synchrotron radiation and laser sources. For example, the new free electron laser (FEL) facilities provide high-energy photon beams with an unprecedented high power within ultra-short pulses, allowing to investigate electron dynamics with a time resolution up to femtosecond.
%This can be used, for the first time ever,
Observing electron exchange during a redox process \cite{DelAnn2013}, melting of charge density waves \cite{SchKir2008} or tuning the electron spin arrangement in a ferromagnet  \cite{Ciu2015} are just some examples of the applications of these sources.

One of the most effective tools to investigate the electronic properties of the materials is the photoelectron spectroscopy (PES), but the effectiveness of this technique in connection with high peak power sources is still questionable.
High peak power implies having a very high number of the photoelectrons in very short time intervals and this may limit the use of this spectroscopic technique with the new pulsed sources.
As a matter of fact, in a standard PES experiment at synchrotron radiation sources, the typical flux is of the order of $\sim$10$^{4}$ photons per pulse,
while the femtosecond FEL pulses are several orders of magnitude more intense. This causes the emission of a large number of photoelectrons interacting with each other through the Coulomb repulsion and  gives rise to the so called space-charge effect \cite{Pie2008}.
This effect may change the  measured energy spectrum by introducing broadening and shift in the spectral feature to an extent that might obscure its physical meaning. For this reason it is of crucial importance to have an effective method capable of simulating this effect in order to find the optimal conditions to minimize it.

J.P. Long, B. S. Itchkawitz and M. N. Kabler (LIK) proposed in 1996 a very simple model to describe the broadening
of photoemission peaks consequent to space-charge effect \cite{LonItc1996}.
The LIK model predicts an energy broadening $\Delta E$ by the
formula:
\begin{equation}
\Delta E \approx \frac{e^2}{2\pi \epsilon_0}\frac{N}{a} \approx 2.9\cdot 10^{-3}\mbox{ eV}\cdot\mu\mbox{m}\frac{N}{a}
\end{equation} \label{eq:LIK}
where $N$ is the number of photoelectrons with charge $-e$ emitted per pulse, $a$ is the radius
of the radiation spot on the sample surface (expressed in $\mu$m) and $\Delta E$ value is given in eV. The energy broadening does not depend on the initial kinetic energy of the electrons.
Despite its strong approximations, and in particular the unrealistic hypothesis that
all the photoelectrons have initially the same energy, the LIK model succeeds
in predicting the order of magnitude of the energy broadening of the PES structures for low excitation energy,
as pointed out by S. Hellmann et al. and Verna et al. \cite{HelRos2009,Verna201614}.
On the other hand, no simple model exists for quantifying the energy shift of the spectral features.
Energy shifts have been reported to
be of the same order of magnitude of the corresponding energy broadening \cite{ZhoWan2005}.

Hellmann et al. \cite{HelRos2009} for the first time used the Treecode software \cite{BarHut1986,treecode}, a program for self-consistent N-body simulation, for calculating the effective coulomb repulsion among single electrons in a cloud during free flight. They obtained that an energy spread and a shift of the spectral features would result as a consequence of the space-charge effect.

This method based on deterministic calculation is effective in predicting values for energy broadening and shift, but the number of calculated trajectories must be equal to the number of electrons in the cloud. This could be a disadvantage when the number of electrons became very large (for examples $\geq$ 20,000 electrons). We would like to underline that Hellmann et al. treated  only free-flying electrons while we want to understand how the space-charge affects the resolution of an electron analyzer.
For evaluating the space-charge effect in an electron analyzer %effect and how it affects the dispersion properties of an electron analyzer
it can be useful a method where the number of calculated trajectories are limited and modifiable. Moreover it is also important to have the possibility to calculate trajectories in the presence of the electrostatic field generated by the electrodes of the spectrometric apparatus.

A different approach to the investigation of the space-charge effect can be performed via the determination of electron trajectories in an electron-optics simulator, such as the SIMION\textsuperscript{\textregistered} software. SIMION\textsuperscript{\textregistered} uses a ray-tracing method based on the finite difference method that solves the Laplace’s equation numerically \cite{manual-SIM,Cas1973,SisZou2008}.
Moreover the SIMION software is one of the most used tools to simulate photoelectron analyzers. In the latest versions of this software package (7.0, 8.0 and 8.1) there is the possibility to introduce the space-charge effect in the  calculation of the trajectories, but the utilization of this tool is subject to various caveats. According to the warning of the SIMION manual \cite{manual-SIM}, the computation methods can give only a qualitative description of the effect of charge repulsion whereas a quantitative analysis of heavily space-charged environments is not guaranteed. One of the limitations is that only interactions between charged particles are taken into account, while the effect of the charge density on the local electric field is not computed \cite{AppDah2005}. Thus it is important to verify the reliability of SIMION in predicting the shift and broadening of spectral feature in PES experiments. As we will see, while we cannot define the limit of this approach in terms of space charge amount, the results of our work suggest that in the regime her explored SIMION is suited to correctly predict charged particles effects.
%Tools that control this effect have been included only in the versions 7.0, 8.0 and 8.1 of the software package. %and it is meaningful to control to  which extent SIMION is able to account for space-charge.
It is important to note that K. Saito et al. \cite{SaiKoi2012} have written a deterministic algorithm to simulate the space-charge in SIMION, but this algorithm is not integrated into the standard version of the software. The SIMBUCA package \cite{VanBec2011} is an alternative software for charged particle optics used in particular to simulate ions in Penning traps and it also allows to calculate the space-charge effects on the particle trajectories \cite{PorMec2015}.

In the present work we will compare the energy broadening  obtained by two different deterministic calculations  considering that at high-energy the stochastic term of the space-charge effect is negligible because the electrons are so fast that the probability to have scattering among the electrons is low \cite{ShoMed2015}.
The two deterministic calculation methods are represented by the SIMION and Treecode softwares and they have been applied to the expansion in free flight of a cloud of mono- and bi-energetic electrons generated by the same high-energy pulse of light. Once confirmed the accuracy of the calculation methods of SIMION in the free flight case, we will present the results of the simulation of a linear time-of-flight (TOF) analyzer optimized for high-energy time-resolved photoemission spectroscopy that takes into account space-charge effects. The paper is organized as it follows: in section \ref{Theofet} the numerical methods at the base of the two programs are presented. In section \ref{freeflight} the results of the two methods for clouds of mono-energetic and bi-energetic electrons in free flight are compared and the simulation for a realistic photoemission from the Cu $2p_{3/2}$ core level is also presented. The section \ref{TOF} is devoted to the space-charge effect in a linear TOF analyzer and  the conclusions are reported in section \ref{concl}.

\section{Methods\label{Theofet}}

In this section the calculation methods at the base of Treecode and SIMION softwares are presented.
Treecode software was initially implemented for the study of astronomical objects
and can be modified to calculate the motion of N particles interacting through Coulomb forces, given
their initial positions and velocities. This software finds an approximate solution of the N-body problem:
in order to calculate the potential acting on the $i$-th electrons for the interaction with the other $N-1$ electrons, the space is recursively divided into cubic cells \cite{BarHut1986}.
For a given cell the position of the center of mass is calculated. If the distance between the center of mass and the $i$-th electron is greater than the side of the cell, all the electrons in the cell are represented
by an effective particle positioned in the center of mass and with a charge equal to $-ne$, where
$n$ is the number of electrons in the cell.
If the side of the cell is greater than the distance between the center
of mass and the $i$-th electron, the
 smaller subcells composing the original
cell are considered.
Following this procedure, the number of mutual interactions
between electrons to be taken into account decreases from $1/2N(N - 1)$ to a
value of the order of $N \ln(N)$ \cite{BarHut1986}.
With this approximation, Treecode calculate all the trajectories for all the interacting electrons in the cloud in vacuum.

The first step in the SIMION way of working is to calculate the electric field in an array defined by the user (discretization of the space), by solving the Laplace's equation ignoring space-charge, and then to determine particle trajectories within this field.
In our work the space charge is taken into account with ion-cloud Coulombic repulsion method \cite{manual-SIM}.
The program does an approximation to estimate Coulomb repulsion of a cloud of electrons,  apportioning the total charge of the electron cloud among the considered trajectories, whose number can be defined by the user.
For example, if we want to simulate an electron cloud of 200,000 electrons, using Treecode we have to consider all the 200,000 trajectories, every trajectory representing a point charge of $-e$.
SIMION simulates a reduced number of trajectories, for example 5000, apportioning the total charge of 200,000 electrons (32 fC) among the considered trajectories \cite{manual-SIM} so that each trajectory contributes to the Coulomb repulsion with a charge of -40 $e$.

For both methods, the initial positions of the electrons are randomly distributed in a circular photon spot on the sample surface. The direction of the initial directions are also randomly distributed in the 2$\pi$ solid angle subtending the upper hemisphere.

\section{Results and discussions}
\subsection{Free flight\label{freeflight}}
\subsubsection{Monoenergetic electron cloud\label{monoc}}

\begin{figure}
\begin{center}
\includegraphics[scale=0.3]{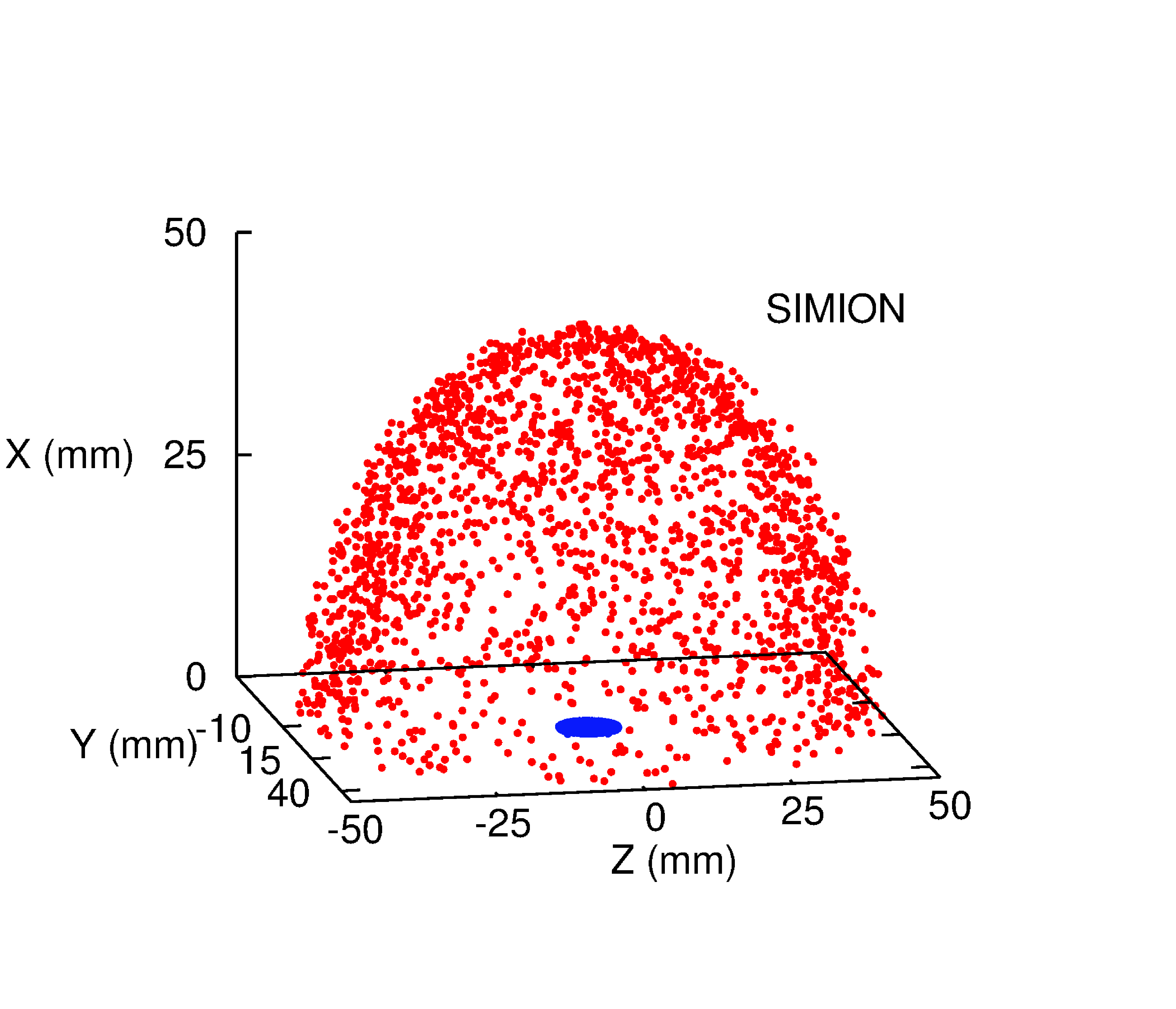}
\end{center}
\caption{Position of the electrons, at time $t$=0 (blue), and at time $t$=0.75 ns after the pulse (red) in vacuum calculated by SIMION; the initial kinetic energy of the electrons is 10260 eV and they originate from a 5-mm radius circular spot. The initial directions are randomly distributed within  angular orientations defined by azimuth (Az=0-180$^{\rm o}$) and elevation (El=0-90$^{\rm o}$) angles. The number of total trajectories is 5000.}\label{xyz}
\end{figure}

In this section we compare and discuss the simulated electron trajectories of electrons in free space expansion obtained by the two methods.
We first considered simulation of an ideal situation in which a single-energy cloud  of photoelectrons starts from a circular spot of 5 mm radius with the same
initial kinetic energy of $\sim$10 keV, in order to perform a first simple comparison of the two methods. The obtained results will be also confronted with the LIK formula.
In Fig. \ref{xyz} the  position of the electrons simulated by SIMION at time $t$=0 and at $t$=0.75 ns after pulse is reported. The $yz$ plane contains the sample surface, where the electrons have origin. The initial directions of the electrons are randomly distributed within a solid angle of $2\pi$ sr.
In order to simulate a large number of electrons emitted at the same time and then to have a good statistics, we consider a particularly large photon spot on the sample surface (5~mm). According to the rough estimations of the LIK formula,  this allows to simulate up to 200,000 electrons, corresponding to a total charge of 32 fC, with an expected moderate energy broadening (about 0.1 eV) \cite{Verna201614}.

\begin{figure}[ht!]
\begin{center}
\includegraphics[scale=0.6]{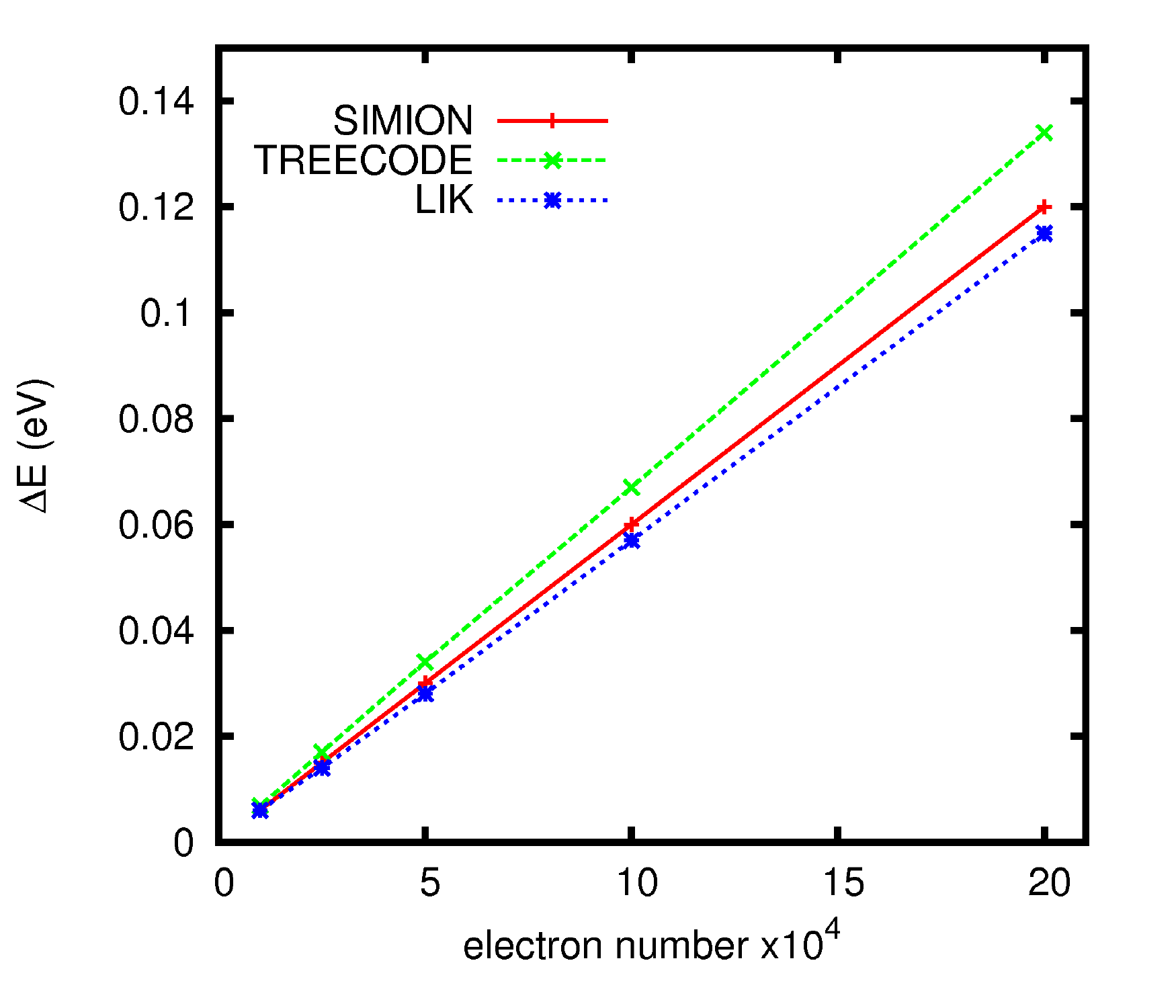}
\end{center}
\caption{Energy spread as a function of the number of photoelectrons in a mono-energetic cloud, for three different calculation: SIMION (solid line), Treecode (dashed line), LIK formula (dotted line). The initial kinetic energy is 10260 eV and the source radius is 5 mm. For the Treecode simulations the number of calculated trajectories corresponds to the number of electrons, while for SIMION the number of trajectories is fixed at 5000 and the total charge of the electrons is apportioned among these trajectories.}
 \label{DE_all}
\end{figure}

In Fig. \ref{DE_all} we compare the  energy spreads as a function of the number of photoelectrons per pulse obtained in the two simulations.
The energy spread $\Delta E$  is defined as the difference between the $5^{th}$ and the $95^{th}$ percentile of the electrons energy distribution.
The results of both methods are in good agreement with  the LIK model (see also Tab. \ref{tab_spread}), that gives good prediction for experiments at lower energy \cite{HelRos2009, Verna201614}.

\begin{figure}
\begin{center}
% a \hskip 4cm b\\
a
\includegraphics[scale=0.25]{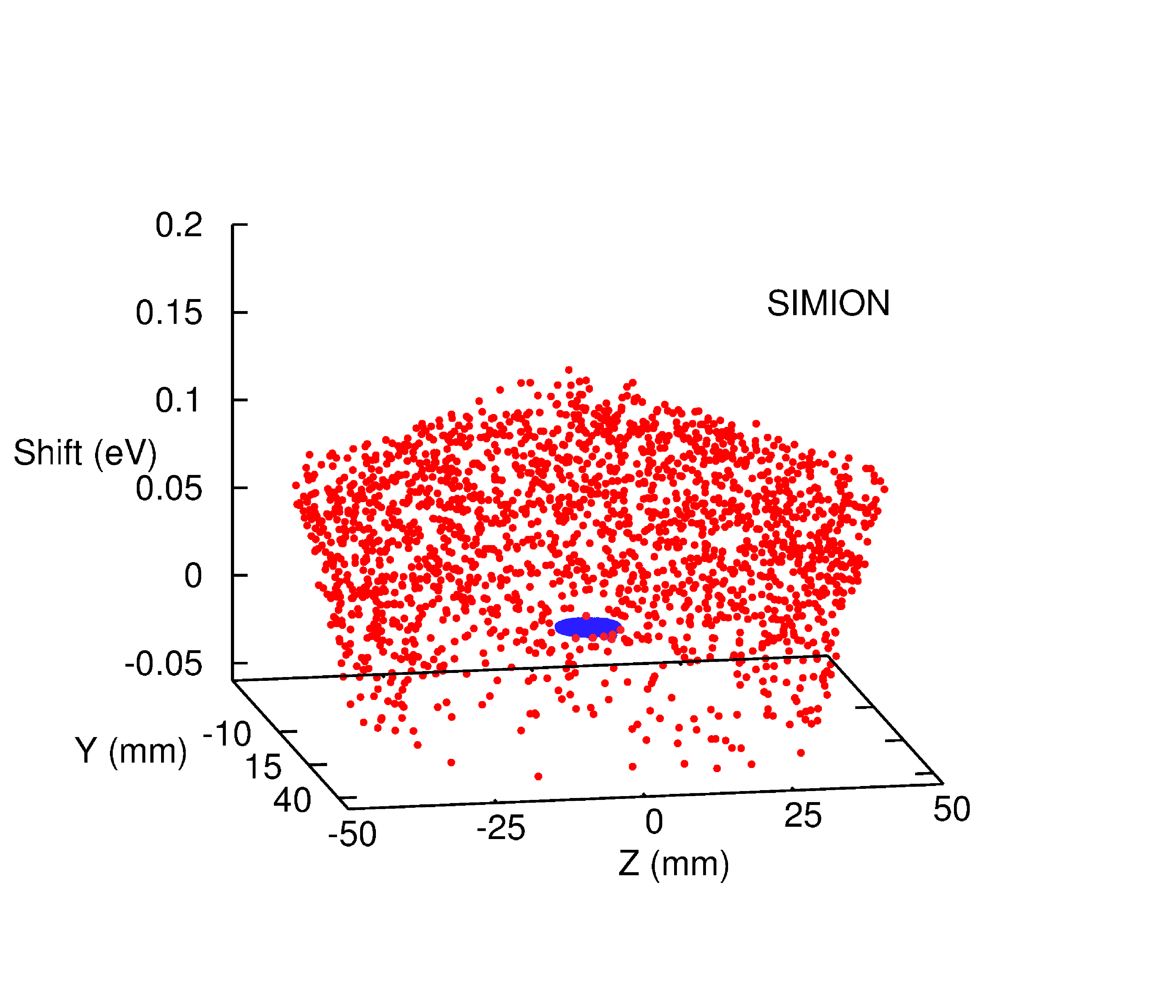}\\
b
\includegraphics[scale=0.25]{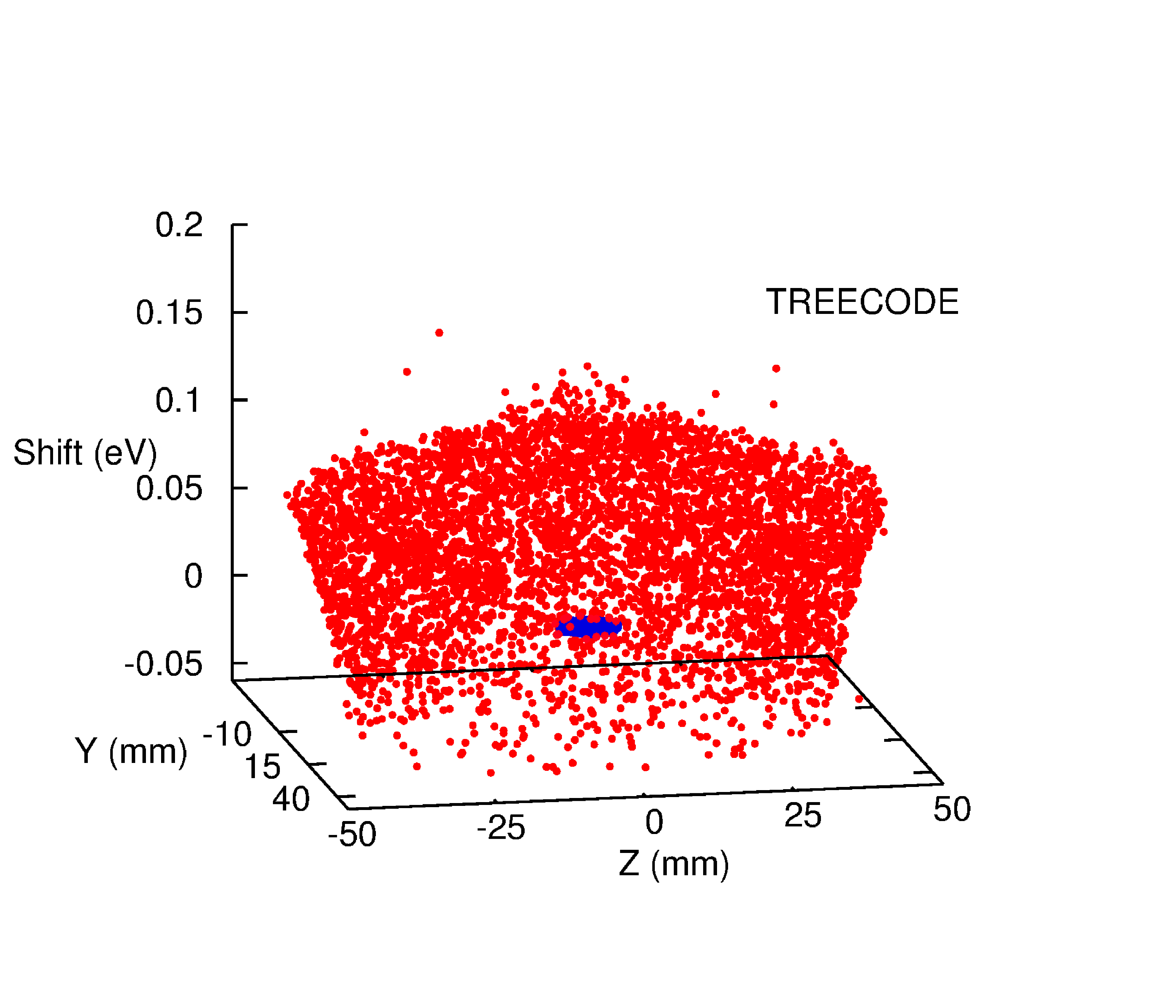}
\end{center}
\caption{Energy shift as a function of the positions in the  plane perpendicular to the electron propagation shown in Fig. \ref{xyz}, at time $t$=0 (blue) and at time $t$=0.75 ns (red) after the pulse. All the electrons have an initial kinetic energy of 10260 eV, the source radius is 5 mm and the total charge is 32 fC. Simulation results from SIMION (5000 trajectories) and Treecode (200,000 trajectories), (a) and (b) respectively.}\label{De-xz}
\end{figure}

Fig. \ref{De-xz} shows the energy shift at $t$=0.75 ns after pulse as a function of the electron $y$ and $z$ positions as obtained by SIMION (a) and Treecode (b).
Qualitatively, the shape-distribution of the energy shifts in the plane of the sample is similar for both calculation methods and clearly departs from a Gaussian function.
The results of the simulations of the space-charge effect in terms of energy spread ($\Delta$E) and average energy shift ($E^{shift}$) for different electrons number and total charge are reported in Tab. \ref{tab_spread}. It has to be noted that the SIMION's results for the energy spread are about 10\% smaller and those for the energy shift about 4\% smaller than what is obtained from Treecode calculations.

\begin{figure}
\begin{center}
% a \hskip 7cm b\\
a
\includegraphics[scale=0.47]{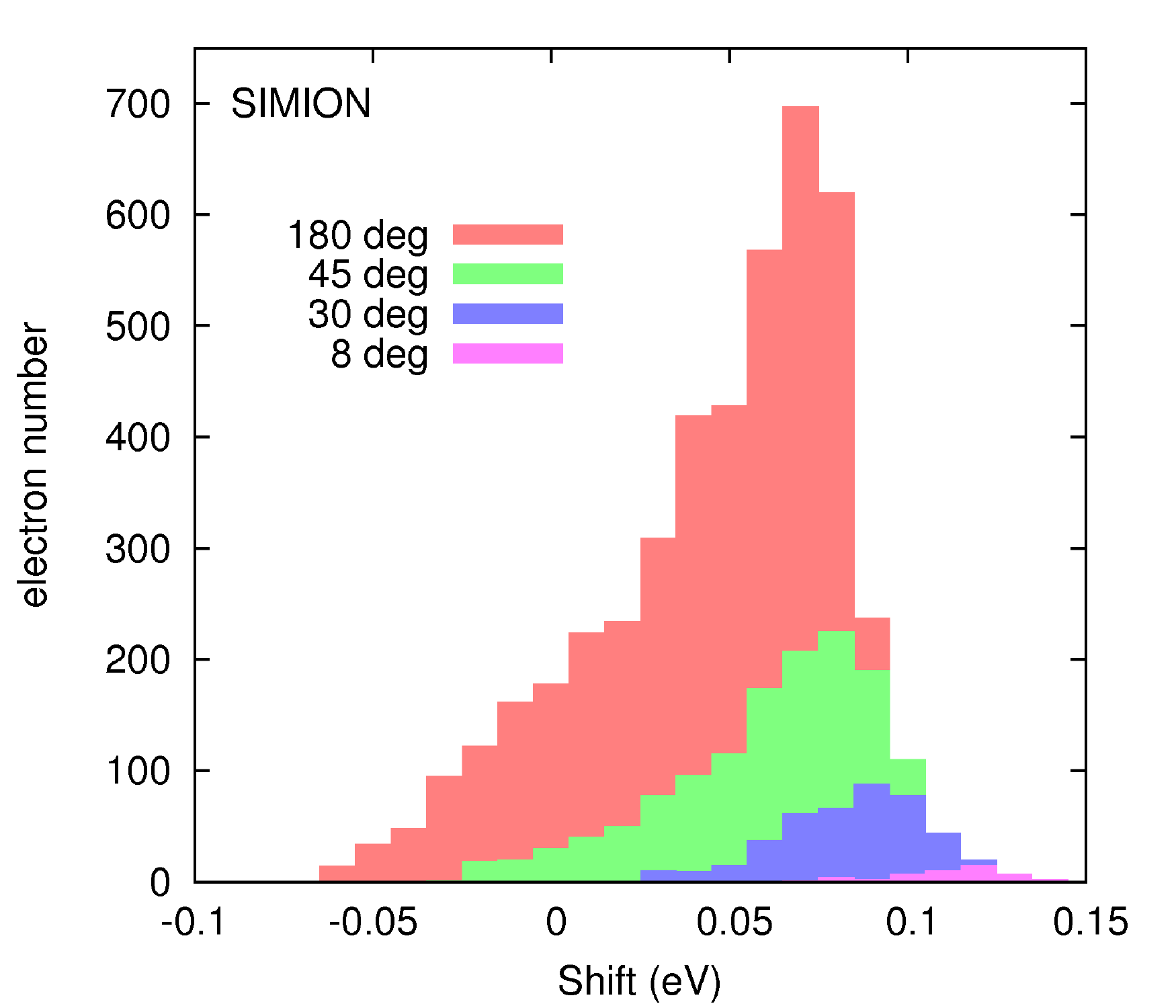}\\
b
\includegraphics[scale=0.47]{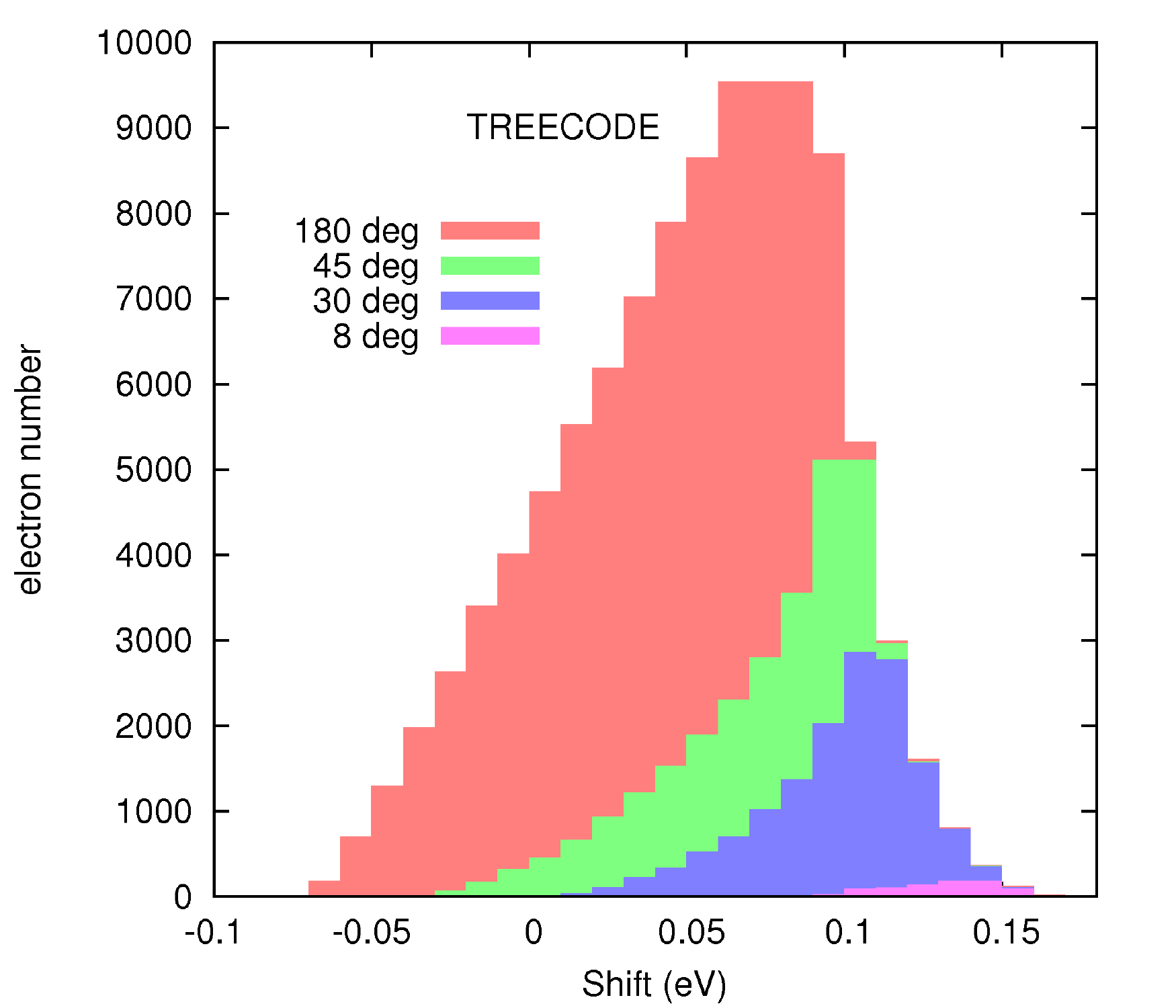}\\
\end{center}
\caption{Distribution of the energy shifts for different accepted polar angles. Initial kinetic energy was set to $E$=10260 eV, radius of the spot is R=5 mm, the azimuthal and polar angles are randomly chosen in the ranges 0$^{\rm o}$--180$^{\rm o}$ and 0$^{\rm o}$--90$^{\rm o}$, respectively. The red histogram includes all the trajectories, in the other histograms we selected only the electrons whose trajectory corresponds to a polar angle smaller than a given value indicated in the label.
The charge repulsion is considered for 200,000 electrons, $Q_{tot}$ =  32 fC. SIMION results are obtained for 5000 trajectories (a) and Treecode results for 200,000 trajectories (b).}\label{Hist}
\end{figure}

We can also investigate the dependence of the energy broadening and shift on the emission angles of the electrons.
We define $\theta$ as the angle between the sample normal (the $x$ axis) and the velocity vector of the electrons.
In Fig. \ref{Hist} and in Tab. \ref{tab_angles} the energy distributions and average energy shift  and energy broadening for accepted $\theta = \pm180^{\rm o}$, $\pm45^{\rm o}$, $\pm30^{\rm o}$, $\pm8^{\rm o}$
 obtained by SIMION (a) and Treecode (b),  are  reported. The center of gravity of the electron energy distributions experiences net positive shift. This is explained by the fact that he electrostatic potential energy of the electrons at time zero, when they are distributed on the disk source, is transformed during the flight into kinetic energy when the particles drift apart. Looking at the Fig. \ref{Hist} and Tab. \ref{tab_angles}, it is possible to note that selecting a smaller solid angle around the normal direction the energy shift increases while the energy spread decreases.
Again this angle dependence predicted by SIMION is in good agreement with the deterministic Treecode simulation.
This effect could be attributed to the fact that in a disk-shaped charge distribution (i.e. the configuration of the photoelectrons at $t$=0) or in a hemisphere-like distribution (i.e. the shape of the photoelectron cloud during the free flight, see Fig. \ref{xyz}) the potential felt by each electron strongly depends on its position.

\subsubsection{Bi-energetic electron cloud: Copper $2p_{3/2}$ case\label{copper}}

In the simulation of a physical spectrum we have to take into account that after the photon excitation of a solid sample at least two processes lead to electron emission in vacuum:
the generation of the primary electrons, due to the photon absorption,  and the emission of the so called secondary electrons,  resulting from inelastic scattering processes within the solid. The latter have a  typical energy less than 20 eV, and they are usually much more numerous than the primary ones emitted at higher kinetic energy.
In this section we consider an electron cloud with two different energies, one representing the electrons of a core-level primary peak and the other representing the large number of secondary electrons. Because in this case we want to simulate a realistic experiment, we consider a smaller spot radius of 500 $\mu$m a more typical value for photoelectron spectroscopy experiments. The total number of electrons is kept fixed at 20,000. In this way we expect from  equation (\ref{eq:LIK}) an energy broadening similar to that obtained for the mono-energetic cloud with a spot radius of 5 mm and 200,000 photoelectrons per pulse.

For example,  in case of excitation with radiation of $h\nu$ = 8000 eV,
the ratio between the number of primary and secondary electrons in the case of the Cu $2p_{3/2}$ core level, with a kinetic energy of 7062.5 eV, can be estimated as 0.0117, where the peak energy for secondary is about 2 eV (details are in Ref. \cite{Verna201614}).

In order to simulate this particular case, the photoelectron cloud with a total charge of 3.2 fC (20,000 electrons) is composed by 231 primary electrons with an initial kinetic energy of 8000 eV (ignoring the binding energy) and by 19769 secondary electrons with initial kinetic energy of 1.8 eV. The results reported below refer only to the primary electrons.
In order to increase the number of trajectories of primary electrons calculated by SIMION (obtaining a better statistics) and to keep a fixed  ratio between primary and secondary electrons, the  trajectories of secondary electrons have a weight of 10, that is one trajectory of secondary electrons carries a charge ten times larger than the charge associated to the trajectory of primary electrons.
In this case the  trajectories calculated by SIMION are 600 for the primary electrons and 5000 for the secondary ones, keeping the exact proportion.

\begin{figure}[ht!]
\begin{center}
%a \hskip 5cm b\\
%\vskip 0.1cm
a
\includegraphics[scale=0.2]{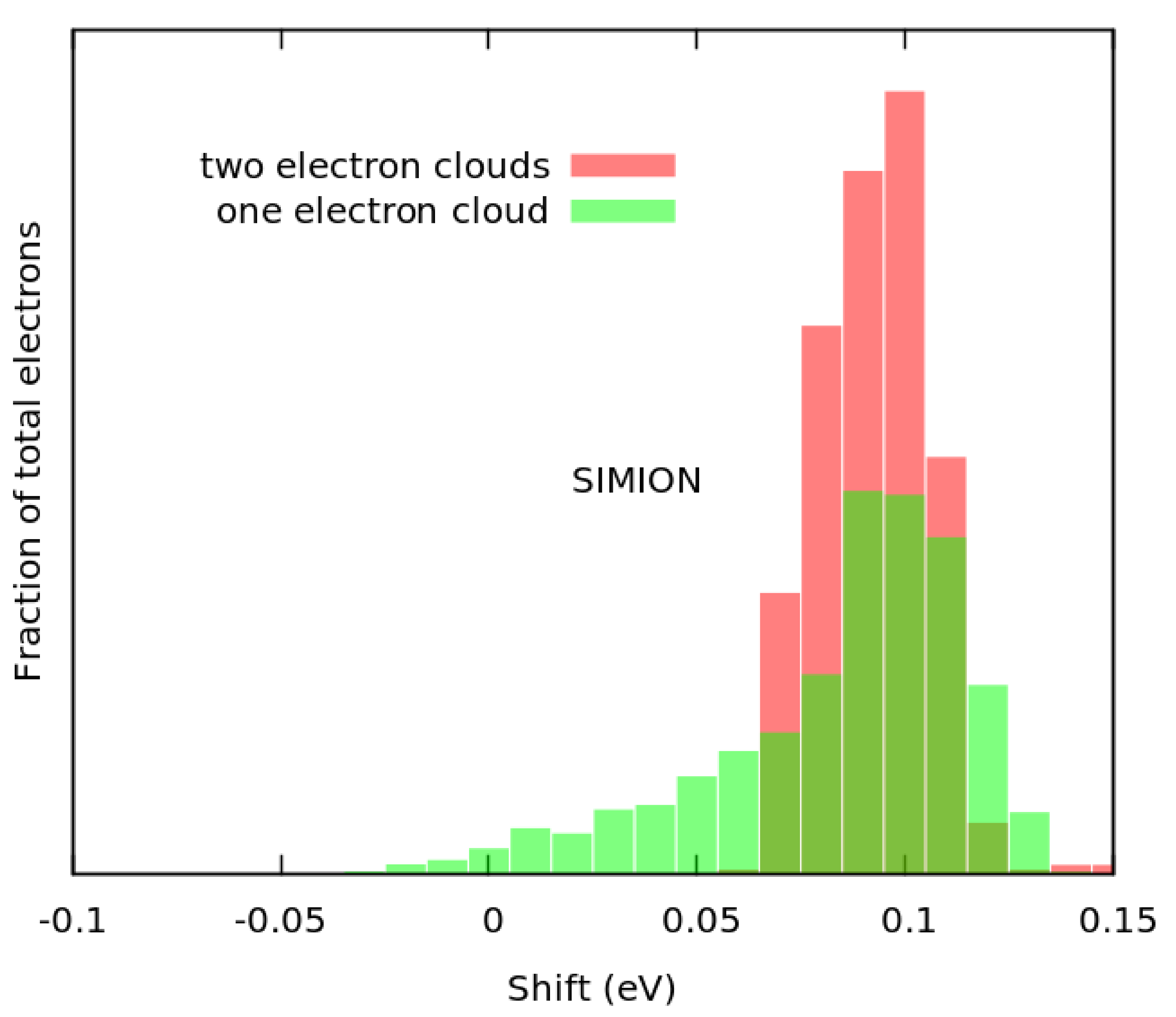}\\
b
\includegraphics[scale=0.2]{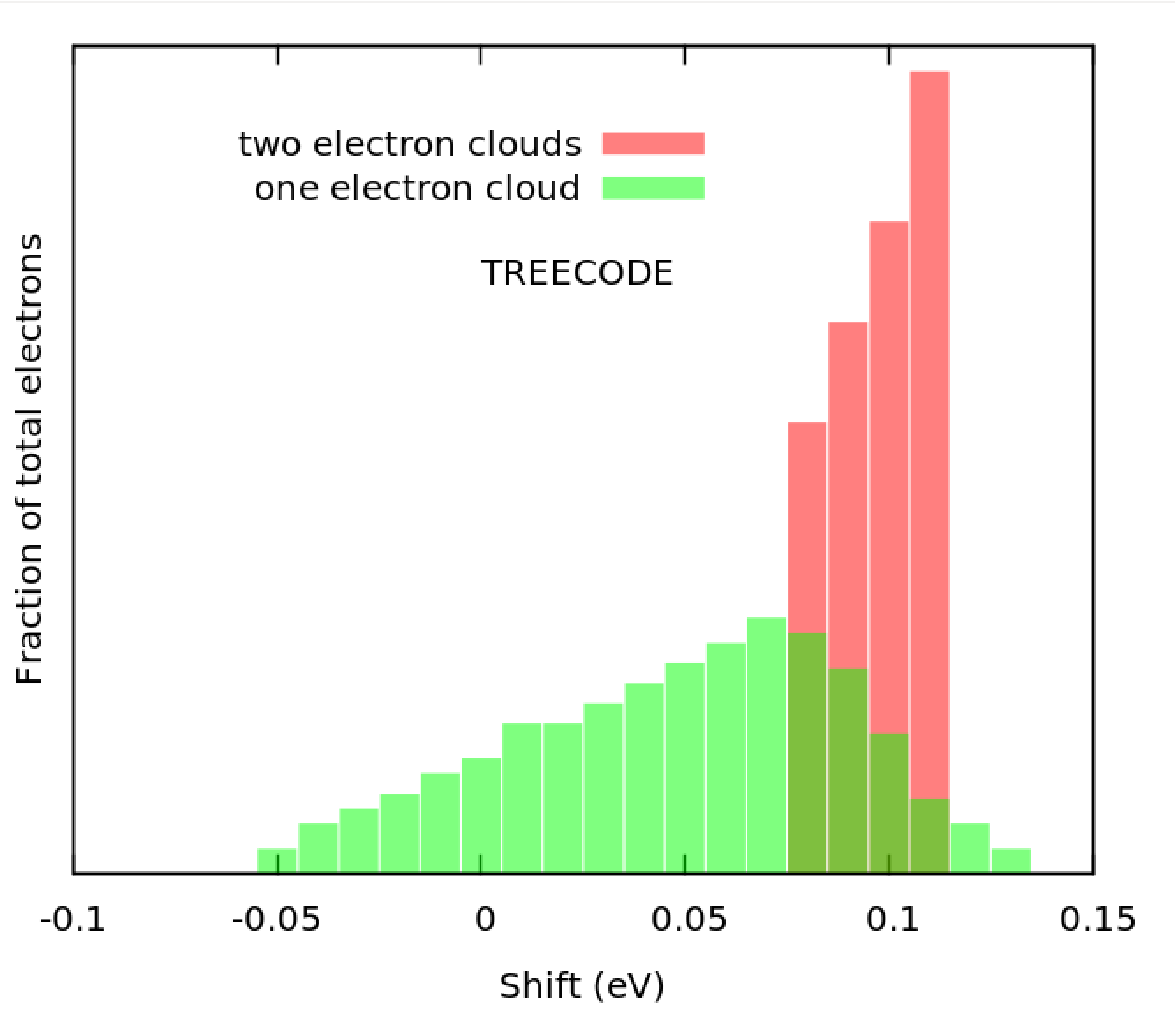}\\
\end{center}
\caption{Red boxes: histogram of the energy shift of primary electrons with the initial kinetic energy of of 8000 eV after the interaction with low-energy ($\sim$2 eV) secondary electrons. This result is compared with the energy spread obtained in a mono-energetic cloud  of primary electrons (green boxes). In both configurations the total charge of the photoelectron cloud is 3.2 fC. Results are shown for SIMION (a) and Treecode (b).}\label{VL-2nubi}
\end{figure}

\begin{figure}[ht!]
\begin{center}
%a \hskip 5cm b\\
a
\includegraphics[scale=0.54]{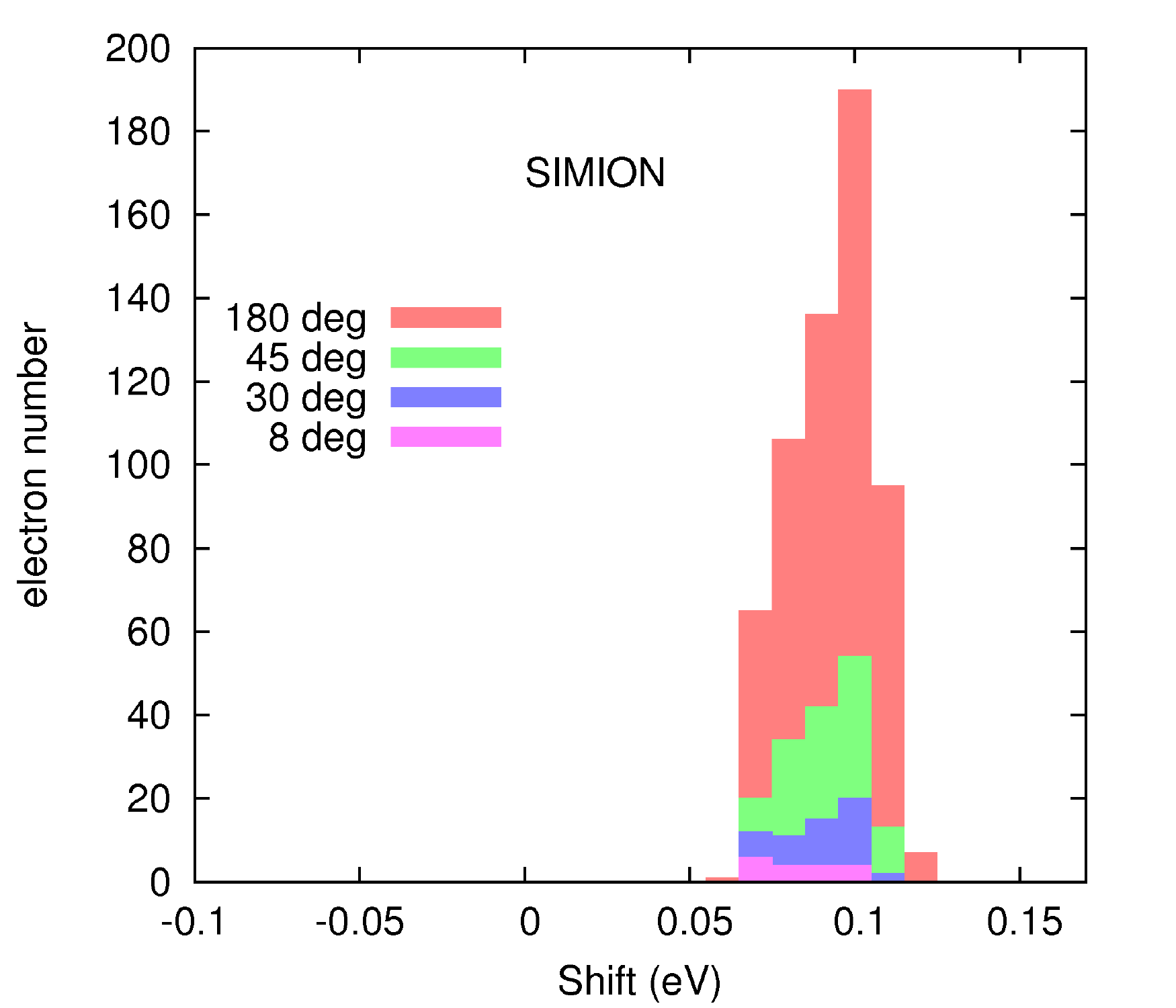}\\
b
\includegraphics[scale=0.54]{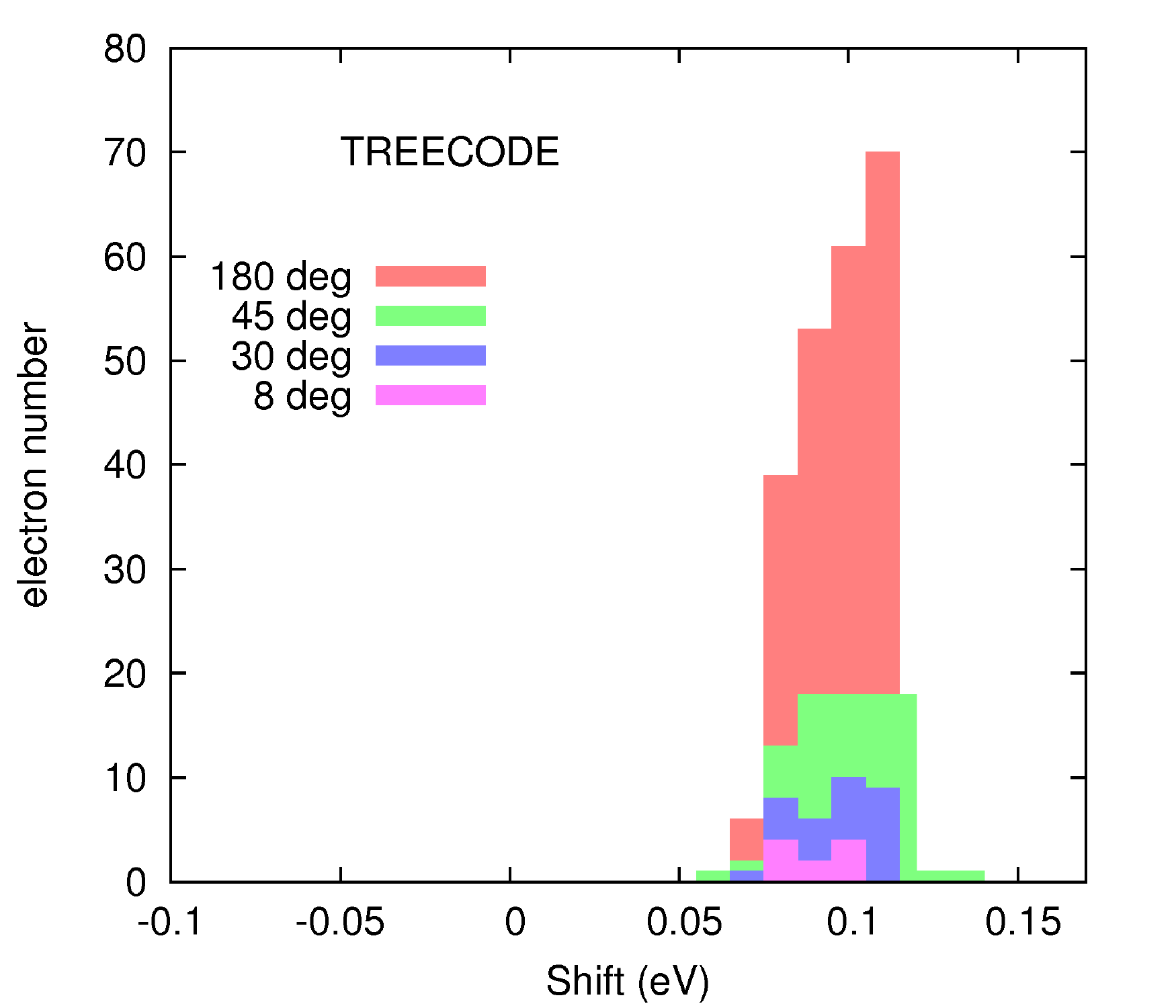}\\
\end{center}
\caption{Histogram of the energy shifts for the primary electrons in a two-energy cloud of photoelectrons as a function of the accepted polar angle. The total charge of the photoelectron cloud is 3.2 fC, the spot radius is 500 $\mu$m and the data are taken 10 ns after the pulse. The results for SIMION (a) and Treecode (b) simulations are reported.}\label{hist_2nubi}
\end{figure}

Figures \ref{VL-2nubi} (a) and (b) show the histograms of the energy shifts obtained for the two simulations, SIMION and Treecode respectively, comparing the results of a monoenergetic cloud  of 20,000 primary electrons with what obtained in the case of two-energy electron cloud.
In the case of two-energy cloud we have to consider that the primary electrons are faster than the secondary ones and thus  the duration of the Coulomb interactions between the two families of electrons is shorter, with a consequent smaller energy spread for primary electrons.
Indeed, Figures \ref{VL-2nubi} (a) and (b), reveal that the effect is taken into account by both simulations. The reduction of the energy spread in the two-energy cloud with respect to the mono-energetic case is $\sim$ 25\%  for SIMION and  $\sim$ 40\%  for Treecode.
The difference in energy shifts between the single- and two-energy cloud is zero for SIMION and 0.05 eV for Treecode.
The secondary electrons  "push" the primary electrons, which should therefore have a greater energy shift compared with the case of the single-energy cloud.
To better investigate this result, in figure \ref{hist_2nubi} the energy shift histogram for different accepted angles is shown. In particular looking at the simulations obtained with SIMION and Treecode (figure \ref{hist_2nubi} (a) and (b)) the energy spread and shift can be considered constant as a function of accepted angle. We remind that in the case of a single electron cloud there was a variation of the energy spread and shift for different accepted angles (see Fig. \ref{Hist}), but these effects are not observed in the case of
the two-energy electron cloud.

\begin{figure}[ht!]
\begin{center}
\includegraphics{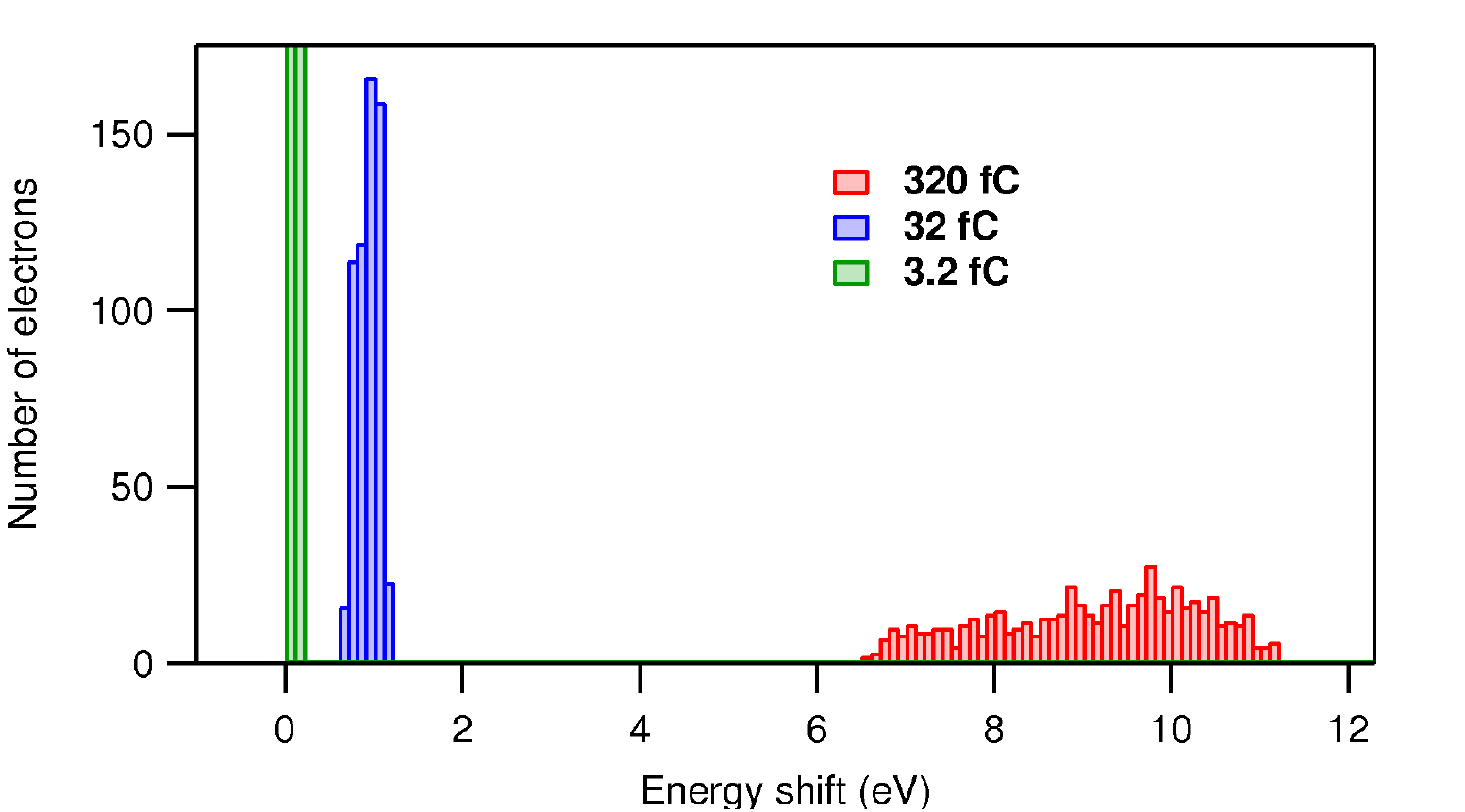}
\end{center}
\caption{Histogram of the energy shifts for the primary electrons in a two-energy cloud of photoelectrons emitted in the free space and having a total charge $Q_{tot}$ of 320 fC (red boxes), 32 fC (blue boxes) and 3.2 fC (green boxes). The spot radius is 500 $\mu$m. Simulation were performed with SIMION and  the data are taken at $t$=1.5 ns. We have verified at this time $E^{shift}$ and $\Delta E$ (see Tab. \ref{tab_highintensity}) have reached their final value. The top of the histogram for $Q_{tot}$=3.2 fC is out of the vertical scale of the graph.}
\label{fig_highintensity}
\end{figure}

For copper the quantum efficiency $\eta$ (i.e. the ratio between the number of the overall photoelectrons and the number of incident photons) results around 0.04 for photon energies around 8 keV. So in order to have 20,000 photoelectrons per shot about 500,000 photon per pulse are requested, implying a strong reduction of the beam intensity available at modern FEL facilities. We want to monitor the space-charge effect when a significantly higher number of photoelectrons are emitted after the same pulse. Using SIMION we repeated the simulation of the two-energy photolectron cloud in free space with a total charge $Q_{tot}$=32 fC ($2\cdot 10^5$ electrons) and  $Q_{tot}$=320 fC ($2\cdot 10^6$ electrons). We do not perform the analogous simulations with Treecode because so a huge number of trajectories would require a very long computation time (some days) or the use of a mainframe computer. With SIMION we still use 600 trajectories of primary electrons (initial energy of 8000 eV) with a weight equal to 1 and 5000 trajectories of secondary electrons (initial energy of 1.8 eV) with a weight equal to 10. Also in this case the total charge is apportioned among the trajectories considering the relative weights so that the ratio 0.0117 between primary and secondary electrons is conserved. In Fig.~\ref{fig_highintensity} we report the histograms of energy distribution of primary electrons at the end of the simulation considering a total charge of photoelectron cloud $Q_{tot}$ of 3.2 fC, 32 fc and 320 fc. The bin-width in the histograms is 0.1 eV. In Tab.~\ref{tab_highintensity} we report the found values of energy shift $E^{shift}_{SIM}$ and broadening $\Delta E_{SIM}$ as a function of the number of photoelectrons and the total charge. Both these quantities increase linearly as a function of the number $N$ of photoelectrons. The simulation with $Q_{tot}$=320 fC ($N$=2$\cdot10^6$) present a photoelectron feature which is tremendously spread ($\Delta E_{SIM}$=3.9 eV). A photoemission cloud with $N$=2$\cdot10^6$ electrons is obtained in an experiment with copper at a photon energy of 8 keV with a pulse of $5 \cdot 10^7$ photons, a value easily achievable in modern FEL beamlines \cite{SchYur2011}. This populated photoemission cloud contains as many as 20000 primary electrons and, in principle, experiments with excellent statistics could be achieved. However, the interaction of the primary electrons with the secondary electrons of the cloud produces an energy broadening of the photoemission peak which is so large to prevent any effective measurement.

%\pagebreak

\subsection{TOF analyzer\label{TOF}}

Taking into account the above considerations and verified the reliability of SIMION in evaluating energy broadening and shift as due to space-charge effect, we can present the simulation results in a linear Time of Flight (TOF) analyzer in order to quantify the energy distortions in a PES experiment at high energy.
The investigated instrument is a linear time-of-flight spectrometer composed by six cylindrical electrodes, the entrance and the exit ones as truncated cones, with a total length of 100 cm and an internal diameter of 10 cm (see Fig. \ref{TOF-TRJ}).
This TOF geometry is optimized for decelerating high energy electrons of about $\sim$10 keV to few eV. The spectrometer resolving power obtained with optimized potential, and neglecting space-charge effect, is of the order of $E/\Delta E=1250$ at $\sim$10 keV. Details are presented elsewhere \cite{lollo}.

\subsubsection{Monoenergetic electron cloud}
\begin{figure}[ht!]
\begin{center}
\includegraphics[scale=0.25]{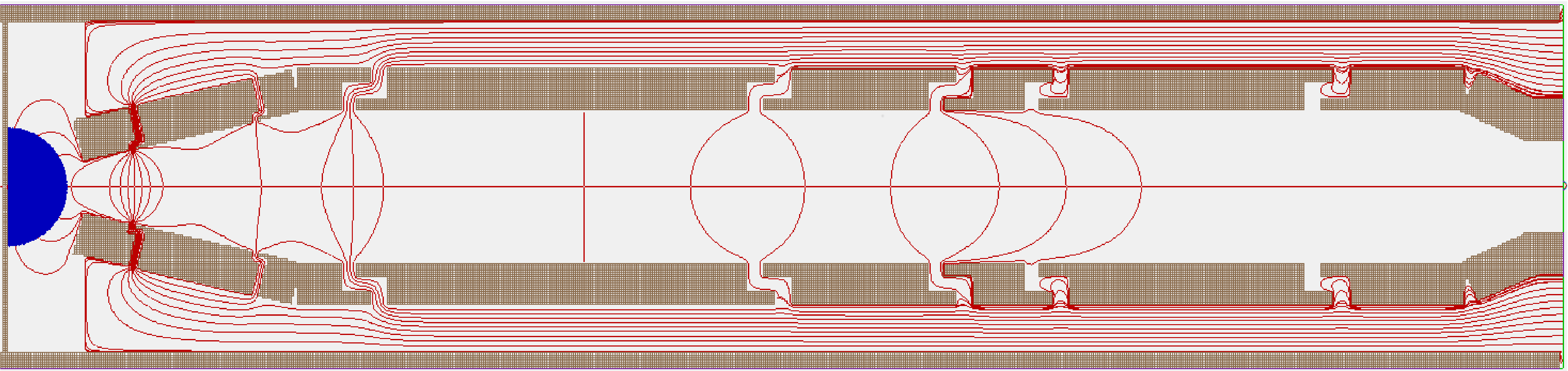}
\includegraphics[scale=0.25]{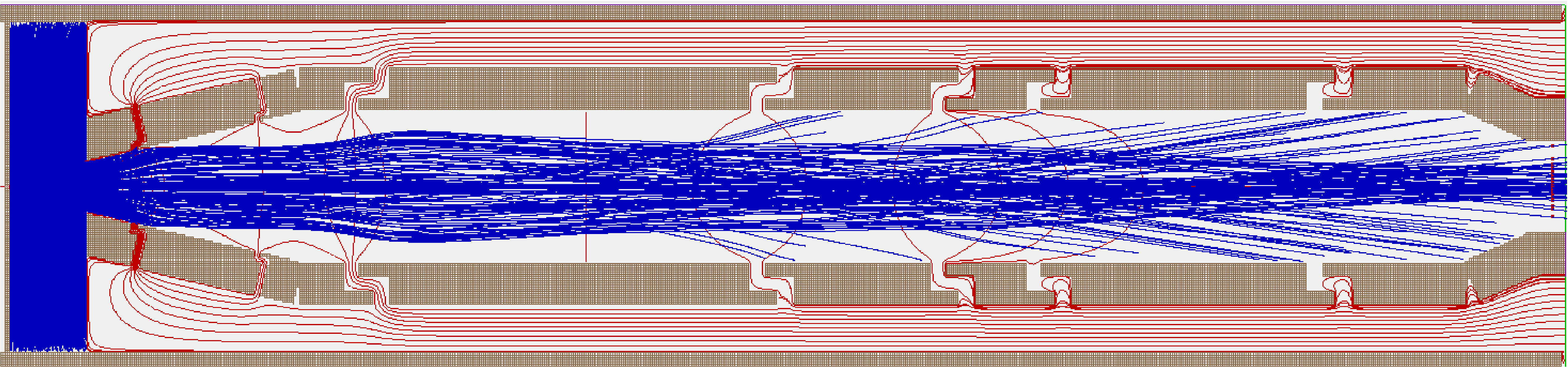}
\end{center}
\caption{5000 trajectories of electrons traveling into a TOF analyzer simulated by SIMION. The initial kinetic energy is 10260 eV and the electrons originate from a spot with 5 mm radius. Upper panel: at $\sim$40 mm from the source and t = 1.3 ns. Bottom panel: electron trajectories up to the impact at the detector at the end of the TOF at t = 65 ns. Red lines represent the equipotential surfaces.}
 \label{TOF-TRJ}
\end{figure}

We first perform simulations for mono-energetic electrons with an initial kinetic energy of 10260 eV and emerging from a circular spot with 5 mm radius with the initial velocities distributed into the upper 2$\pi$ solid angle. Fig. \ref{TOF-TRJ} shows 5000 trajectories into the TOF, the electron cloud possessing the total charge of 200,000 electrons. The trajectories  are depicted from $t=0$ to $t=1.3$ ns (upper panel) and from $t$=0 to $t=65$ ns (bottom panel). Due to the large source dimension and the wide emission solid angle, only about  $1\%$ of the emitted electrons  arrive to the end of the analyzer (Fig. \ref{TOF-TRJ} bottom panel). In particular, the accepted solid angle results to be about 80 msr.

\begin{figure}[ht!]
\begin{center}
%a \hskip 5cm b\\
%\includegraphics[scale=0.3]{../../img/grph-SIM/ok/Hist-320pC_TOF_pot5000trj.ps}
\includegraphics[scale=0.5]{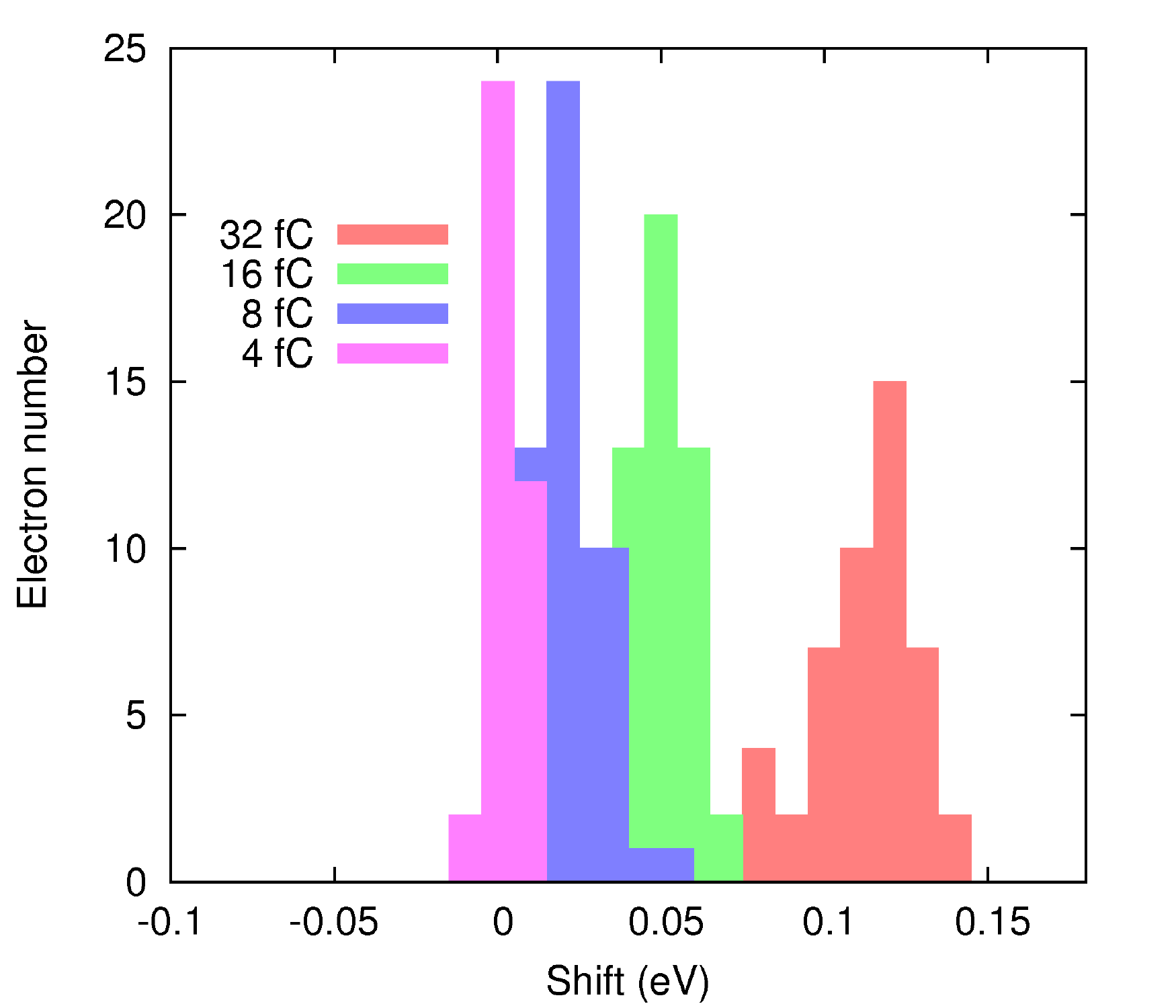}
\end{center}
\caption{Energy shift distribution of electrons with initial kinetic energy of 10260 eV and emerging from a spot of 5 mm  radius at the end of the linear TOF analyzer, for different values of the total charge of the photoelectron cloud (4--32 fC), as indicated in the figure key. Calculations were carried out by SIMION and the total charge is apportioned among a fixed number of 5000 trajectories.}
 \label{elect-plane_zy}
\end{figure}

In Tab. \ref{tab_spread-tof} we show the space-charge effect in terms of energy spread and shift at the end of the linear TOF  for different numbers of photoemitted electrons and in Fig. \ref{elect-plane_zy} the corresponding histograms of energy are shown. As expected, the energy shift and spread increase with the total charge of the electrons. In these conditions the space-charge effect becomes negligible for 25000 electrons per cloud as also shown in Tab. \ref{tab_spread-tof}.

\subsubsection{Bi-energetic electron cloud: Copper 2$p_{3/2}$ case}

\begin{figure}[ht!]
\begin{center}
\includegraphics[scale=0.4]{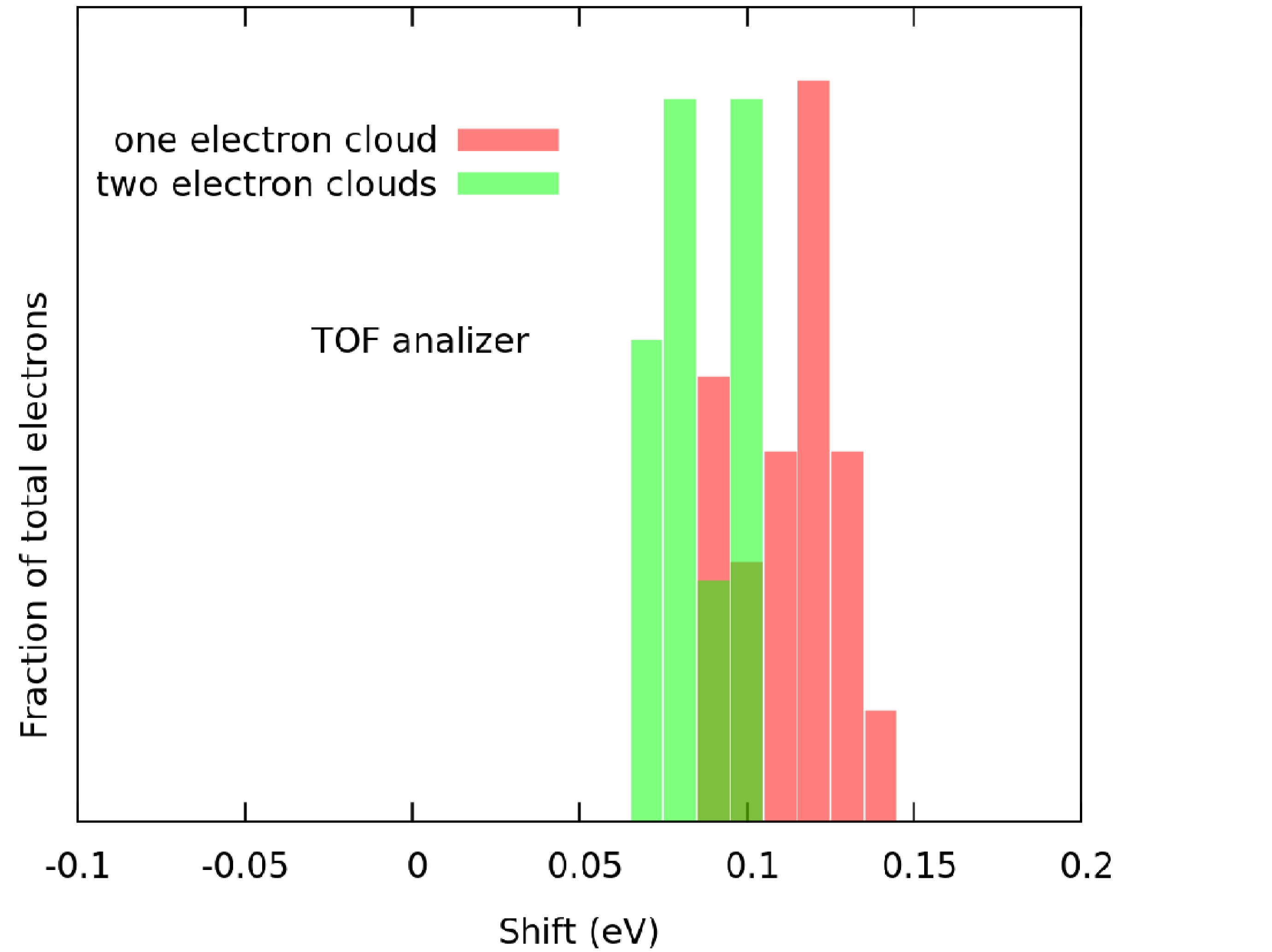}
\end{center}
\caption{Histogram of energy shifts at the end of a TOF analyzer for primary electrons (8000 eV) after an interaction with low-energy ($\sim$2 eV) secondary electrons (red boxes). This result is compared with the energy spread obtained by a single-energy cloud  of primary electrons (green boxes). The total charge is 3.2 fC and the spot radius is 500 $\mu$m}\label{TOF-2nubi}
\end{figure}

In this case we have simulated the two clouds of electrons  described in details in section \ref{copper} during their propagation into a TOF analyzer. Again, this case mimics a realistic experiment of photoemission from the Cu $2p_{3/2}$ core level. In Fig. \ref{TOF-2nubi} the obtained histograms of the energy shift for the single-energy and for the two-energies cloud are compared.
It is possible to note that, as expected, for the two-energy cloud the energy spread is lower than in the case of mono-energetic cloud but the same does not occur fo the energy shift. Looking at the results reported in Tab. \ref{tab_2nubi-TOF} the energy shift obtained for mono-energetic cloud of electrons is larger by $\sim$33\% than the value obtained for the two-energy cloud. This is due to the accepted angular dependence of the single-energy cloud, in fact only electrons within an accepted angle of $\pm$ 8$^{\rm o}$ can enter in the TOF analyzer here presented. As we have seen in the case of flight through free space, mono-energetic electrons emitted into a small accepted cone experience a larger energy shift (see Fig.~\ref{Hist} and Tab.~\ref{tab_angles}. This larger positive shift is absent for primary electrons in the two-energy cloud (see Fig.~\ref{hist_2nubi}) and this explains the difference in average energy with the monoenergetic case.

%\pagebreak

\section{Conclusions\label{concl}}
In this work we have compared two deterministic methods for calculating the space-charge effect: the first based on the SIMION software and the second exploiting the Treecode algorithm.
From this comparison we want to control the effectiveness of the deterministic method used by SIMION for predicting this effect.
The use of SIMION mainly presents two important advantages: 1) the calculation with a limited number of trajectories (suitably chosen by the user) representing a cloud with a much larger number of photoelectrons; 2) the chance to simulate the trajectories in presence of an external electric (and in perspective magnetic) field and then the possibility to calculate the energy distortions of spectra in an electron analyzer.
The results obtained by the two methods of simulation are in good agreement for the different treated cases  of a photoelectron cloud freely expanding in vacuum. In particular, we have simulated a single-energy cloud and a two-energy cloud composed by primary and secondary electrons closely resembling the realistic case of photoemission from the Cu 2$p_{3/2}$ core level in free flight.
We can conclude that the method used by SIMION is reliable and with this tool we have extended the in-vacuum calculations to the simulations of photoelectrons flying across a time-of-flight linear analyzer provided with decelarating electrostatic lenses \cite{lollo}.

Moreover we have found that for a monoenergetic electron cloud the energy spread depends on the final position and the emission angle of the photoelectrons.
This could be very useful not only to diminish the distortions due to space-charge effect without decreasing the number of photoelectrons per pulse, but also to correct such distortions. It has to be underlined that this is valid only for the monoenergetic case while it appears to be absent for the small number of high-energy primary electrons in the two-energy cloud. Moreover we can conclude that a linear TOF is suitable for a PES experiment with a high-energy FEL source.

\section*{Acknowledgments}
This work was partially supported by the ULTRASPIN and EX-PRO-REL
Projects in kind (PIK), funded by Ministero dell'Istruzione, dell'Universit\`a e
della Ricerca (MIUR) through Elettra Sincrotrone Trieste. G. G. is grateful to
 the PIK project EX-PRO-REL for financing her postdoctoral fellowship. V. L.
acknowledges the PIK project ULTRASPIN for granting his fellowship. A. V.
is thankful to Regione Lazio and CRUL for financial support.
% If in two-column mode, this environment will change to single-column format so that long equations can be displayed.
%\begin{widetext}
%$$\mbox{put long equation here}$$
%\end{widetext}

% Figures should be put into the text as floats.
% Use the graphics or graphicx packages (distributed with LaTeX2e).
% Put the label that you will use with \ref{} command in the braces of the \label{} command.
%
% \begin{figure}
% \end{figure}

% \begin{table}
% \end{tabular}
% \end{table}

%\begin{acknowledgments}
% Put your acknowledgments here.
%\end{acknowledgments}

%\bibliography{bib_GGreco_modified_AV_FO}
%\bibliographystyle{elsarticle-num}

\begin{table}[ht!]
\begin{center}
 \begin{tabular}{d{3}d{3}ccd{3}d{3}c}
\hline
\hline
\multicolumn{1}{l}{Free flight}\\
\hline
\multicolumn{1}{c@{\hspace{-0.1 cm}}}{$N$} & \multicolumn{1}{c@{\hspace{0.6 cm}}}{$Q_{tot}$}    &  $\Delta$E$_{SIM}$ & $E^{shift}_{SIM}$ & \multicolumn{1}{c}{$\Delta$E$_{TRE}$} & \multicolumn{1}{c}{$E^{shift}_{TRE}$} & $\Delta$E$_{LIK}$\\
\multicolumn{1}{c@{\hspace{-0.1 cm}}}{$\times$10$^4$} &  \multicolumn{1}{c@{\hspace{0.6 cm}}}{fC}  & eV & eV & \multicolumn{1}{c}{eV} & \multicolumn{1}{c}{eV} & eV  \\
\hline
20 & 32 &0.120&0.05&0.134&0.048&0.115\\
10 & 16 &0.060&0.03&0.067&0.024&0.057\\
5.0 & 8 &0.030&0.01&0.034&0.012&0.028\\
2.5 & 4 &0.015&0.01&0.017&0.0062&0.014\\
1.0 & 1.6 &0.006&0.00&0.0068&0.0024&0.006\\
\hline
\hline
\end{tabular}
\caption{Space-charge effect in terms of energy spread ($\Delta E$) and shift ($E^{shift}$) for different number of electrons $N$ and total charge $Q_{tot}$ in a mono-energetic cloud. Data are acquired at time $t\sim$ 10 ns after light pulse, when $\Delta E$ and $E^{shift}$ have reached their final value.
 $\Delta$E$_{M}$ and $E^{shift}_{M}$ are the energy spread and shift calculated by SIMION (M = SIM) and Treecode (M = TRE), while $\Delta$E$_{LIK}$ is the energy spread calculated by the LIK formula. The initial kinetic energy is 10260 eV for all the electrons with a source radius of 5 mm. For Treecode the number of trajectories corresponds to the real number of the electrons while for SIMION the number of trajectories is fixed at 5000 and the total charge $Q_{tot}$ is apportioned among them so that the simulated repulsive charge is the same for the two methods.
}\label{tab_spread}
\end{center}
\end{table}

\begin{table}[ht!]
\begin{center}
 \begin{tabular}{rrrcccc}
\hline
\hline
Free flight\\
\hline
$\theta$ & N$_{SIM}$ & N$_{TRE}$&  $\Delta$E$_{SIM}$ & $E^{shift}_{SIM}$ &  $\Delta$E$_{TRE}$ & $E^{shift}_{TRE}$ \\
deg. & &&eV & eV & eV & eV  \\
\hline
180 &5000&200000& 0.120&0.060&0.150&0.07\\
45  &1700&58306&0.100&0.070&0.080&0.10\\
30 &530&26619&0.070&0.085&0.074&0.11\\
8 &50&1469&0.020&0.120&0.035&0.14\\
\hline
\hline
\end{tabular}
\caption{Space-charge effects in terms of energy broadening ($\Delta$E) and shift ($E^{shift}$ for different accepted angles data are acquired after $\sim$ 10 ns.
$\theta$ is the opening angle of the accepted cone in degrees; N$_{M}$ (M = SIM or TRE) are the effective number of trajectories selected by the correspondent solid angle; $\Delta$E$_{M}$, $E^{shift}_{M}$ (M = SIM or TRE), are the energy spread and shift.
The initial kinetic energy of the photoelectrons  is 10260 eV and the source radius is 5 mm. The total charge $Q_{tot}$ is 32 fC, corresponding to 200,000 electrons. The number of calculated trajectories is 5000 for SIMION and 200,000 for Treecode.
}\label{tab_angles}
\end{center}
\end{table}


\begin{table}[ht!]
\begin{center}
\begin{tabular}{d{3}d{3}d{3}d{3}}
\hline
\hline
\multicolumn{1}{l}{Free flight}\\
\hline
\multicolumn{1}{c}{$N$}  & \multicolumn{1}{c}{$Q_{tot}$} & \multicolumn{1}{c}{$E^{shift}_{SIM}$} & \multicolumn{1}{c}{$\Delta E_{SIM}$}\\
\multicolumn{1}{c}{$\times 10^4$}   & \multicolumn{1}{c}{fC}        & \multicolumn{1}{c}{eV}                & \multicolumn{1}{c}{eV} \\
\hline
200             & 320       & 3.9               & 9.1\\
20              & 32        & 0.38              & 0.92\\
2               & 3.2       & 0.038             & 0.093\\
\hline
\hline
\end{tabular}
\end{center}
\caption{Average energy shift $E^{shift}_{SIM}$ and energy broadening $\Delta E_{SIM}$ simulated by SIMION for a two-energy photoemission cloud flying in free space considering the indicated values of number of electrons $N$ and total charge $Q_{tot}$. The data are taken at $t$=1.5 ns, when $E^{shift}_{SIM}$ and $\Delta E_{SIM}$ have reached their final value.}
\label{tab_highintensity}
\end{table}

\begin{table}[ht!]
\begin{center}
 \begin{tabular}{rrcc}
\hline
\hline
linear TOF\\
\hline
$N$ & $Q_{tot}$    &  $\Delta$E$_{SIM}$ & $E^{shift}_{SIM}$ \\
$\times$10$^4$ &  fC  & eV & eV  \\
\hline
20 & 32 &0.05&0.12\\
10 & 16 &0.04&0.08\\
5.0 & 8 &0.02&0.05\\
2.5 & 4 &0.00&0.04\\
\hline
\hline
\end{tabular}
\caption{Space-charge effect of electrons with initial kinetic energy of 10260 eV and spot radius of 5 mm, in terms of energy spread ($\Delta$E) and shift $E^{shift}$ for different number of electrons and total charge. Simulations were carried out with SIMION using a fixed number of 5000 trajectories.}\label{tab_spread-tof}
\end{center}
\end{table}

\begin{table}[ht!]
\begin{center}
 \begin{tabular}{lcc}
\hline
\hline
    &  $\Delta$E$_{SIM}$ & Shift$_{SIM}$ \\
  & eV & eV   \\
\hline
\hline
TOF - one-energy cloud &0.06&0.12\\
TOF - two-energies cloud &0.04&0.08\\
\hline
\hline
\end{tabular}
\caption{Space-charge effect in a TOF analyzer in terms of energy spread ($\Delta$E) and shift ($E^{shift}$) considering a two-energy cloud of primary (initial kinetic energy 8000 eV) and  secondary electrons (initial energy 1.8 eV) and a cloud of primary electrons (initial kinetic energy 8000 eV). The total number of electrons is 20,000 (3.2 fC) and the ratio between primary and secondary electrons is 0.0117. The spot radius is 500 $\mu$m. Data are acquired after $\sim$ 10 ns of flight.
}\label{tab_2nubi-TOF}
\end{center}
\end{table}

\end{document}